\documentstyle[trans]{rspublic}

\newcommand{\beet}{\begin{equation*}}
\newcommand{\eeet}{\end{equation*}}
\newcommand{\beaet}{\begin{eqnarray*}}
\newcommand{\eeaet}{\end{eqnarray*}}
\newcommand{\bfig}{\begin{figure}}
\newcommand{\efig}{\end{figure}}
\newcommand{\bc}{\begin{center}}
\newcommand{\ec}{\end{center}}

\newcommand{\szz}{\mbox{$\sigma_{zz}$}}

\newcommand{\srr}{\mbox{$\sigma_{rr}$}}

\newcommand{\szr}{\mbox{$\sigma_{zr}$}}
\newcommand{\srz}{\mbox{$\sigma_{rz}$}}

\newcommand{\scc}{\mbox{$\sigma_{\chi \chi}$}}

\newcommand{\src}{\mbox{$\sigma_{r \chi}$}}

\newcommand{\szc}{\mbox{$\sigma_{z \chi}$}}

\begin{document}

\title[Stresses in Poured Sand]
{Development of Stresses in Cohesionless Poured Sand}

\author[Cates,Wittmer,Bouchaud,Claudin]
{M.~E.~Cates $^1$, J.~P.~Wittmer $^1$, J.-P.~Bouchaud $^2$, P.~Claudin $^2$}
\affiliation{
$^1$ Dept. of Physics and Astronomy, University of Edinburgh\\
JCMB King's Buildings, Mayfield Road, Edinburgh EH9 3JZ, UK.\\
$^2$ Service de Physique de l'Etat Condens\'e, CEA\\
Ormes des Merisiers, 91191 Gif-sur-Yvette, Cedex France. }

\maketitle
\label{firstpage}

\begin{abstract}
The pressure distribution beneath a conical sandpile,
created by pouring sand from a point source onto a rough rigid support,
shows a pronounced minimum below the apex (`the dip').  Recent work of
the authors has attempted to explain this phenomenon by invoking local
rules for stress propagation that depend on the local geometry, and hence
on the construction history, of the medium. We discuss the fundamental
difference between such approaches, which lead to hyperbolic
differential equations, and elastoplastic models, for which the
equations are elliptic within any elastic zones present. In the hyperbolic
case, the stress distribution at the base of a wedge or cone (of given
construction history), on a rough rigid support, is uniquely determined by
the body forces and the boundary condition at the free (upper) surface. In
simple elastoplastic treatments one must in addition specify, at the base of
the pile, a displacement field (or some equivalent data). This displacement
field appears to be either ill-defined, or defined relative to a reference
state whose physical existence is in doubt. Insofar as their predictions
depend on physical factors unknown and outside experimental control,
such elastoplastic models predict that the observations should be
intrinsically irreproducible. This view is not easily
reconciled with the existing experimental data on conical sandpiles, 
which we briefly review. Our hyperbolic models are based instead on a
physical picture of the material, in which
(a) the load is supported
by a skeletal network of force chains (``stress paths") whose
geometry depends on construction history;
(b) this network is `fragile' or marginally stable, in a sense that we define.
Although perhaps oversimplified, these assumptions may lie closer to the
true physics of poured cohesionless grains than do those of conventional
elastoplasticity. We point out that our hyperbolic models can nonetheless be
reconciled with elastoplastic ideas by taking the limit of an extremely
anisotropic yield condition.

\end{abstract}

\newpage
\section{Introduction}
\label{sec:Introduction}
Recently, a new strategy for the modelling
of stress propagation in static cohesionless granular media was developed
(Bouchaud {\em et al.} 1995; Wittmer {\em et al.} 1996, $1997a,b$).
The medium is viewed as an assembly of rigid particles held up by friction.
The static indeterminacy of frictional forces within the assembly is
circumvented by the assumption of certain local {\em constitutive
relations} (c.r.'s) among components of the stress tensor.\footnote{
In solid mechanics
the term `constitutive relation' normally refers to a material-dependent
equation relating stress and strain. In fluid mechanics one has instead
equations relating stress and (in the general case of a viscoelastic
fluid) strain-rate history. Our models of granular media entail
equations relating stress components to one another,
in a manner that we take to depend on the construction history of the
material. Clearly such equations describe constitutive properties of
the medium: they relate its state of stress to other
discernable features of its physical state. We see no alternative
to the term `constitutive relations' for such equations. The same equations
could, of course, be obeyed by some solutions of models whose
constitutive definition was quite different; in that context they would not
be c.r.'s. }
These are assumed to encode the network of contacts in the granular
packing geometry; they therefore depend explicitly on the way in
which the medium was made -- its {\em construction history}. The
task is then to postulate and/or justify physically suitable
c.r.'s among stresses, of which only one (the {\em primary} c.r.)
is required for systems with two dimensional symmetry, such as a
wedge of sand;  for a three dimensional symmetric assembly
(the conical sandpile) a secondary c.r. is also needed.

Among the primary constitutive relations of Wittmer {\em et al.} \mbox{(1996,
$1997a$)} are a certain class  (called the `oriented stress linearity' or
{\sc OSL} models) which have simplifying features. Indeed, in
two-dimensional geometries these combine with the stress continuity equation
to give a wave equation for stress propagation, in which the horizontal
and vertical directions play the role of spatial and temporal coordinates
respectively (Bouchaud {\em et al.} 1995).
A distinguishing feature of the {\sc OSL} models is that the
{\em characteristic rays for stress propagation}
(analagous to light or sound rays in ordinary wave propagation)
are then fixed by the construction history: they do not
change direction under subsequent reversible loading.
(Irreversible loadings, which can in these models be infinitesimal,
are discussed in Section~\ref{sec:Fragile} below.)
As discussed by Bouchaud {\em et al.} (1998),
the characteristics of the differential equation can be viewed as
representing, in the continuum, the mean behaviour of `force chains' or 
`stress paths' in the material
(Dantu 1967; Liu {\em et al.} 1995; Thornton \& Sun 1994).

Of the {\sc OSL} models, a particularly appealing member,
with special symmetry properties, is called the
`fixed principal axes' ({\sc FPA}) model. This has the additional
property that the characteristics everywhere coincide in
orientation with the principal axes of the stress tensor.
The {\sc FPA} model therefore supposes that these principal axes
have an orientation fixed at the time of burial.
\footnote{
Gudehus (1974) has previously used a related idea, that
the principal axes should be locally specified as inputs to
the stress continuity equations. This he employs as a calculation
method for generating a variety of stress distributions based on
`gutem statischen Gef\"uhl'.
}
This is arguably the simplest possible choice for a
history-dependent c.r. among stresses.  For the case of a sandpile in which
grains are deposited by surface avalanches, which we presume to apply for a
conical pile constructed from a point source (though see
Section~\ref{sec:Experiment}\ref{subsec:plasticone} below),
the orientation of the major axis at burial is
constant, and known from the fact that the free surface in such a pile
must be a yield surface. The resulting constitutive equation among
stresses, for the sandpile geometry, then has a singularity at the centre
of a cone or wedge; this is physically admissible since the centreline
separates material which has avalanched to the left from material which
has avalanched to the right. This singularity leads to an `arching'
effect, as previously invoked to explain the stress-dip by
Edwards \& Oakeshott (1987) and others
(Trollope 1968; Trollope \& Burman 1980).

The {\sc OSL} models were developed to explain experimental
data on the stress distribution beneath a conical sandpile, built by
surface avalanches of sand, poured from a point source onto a rough,
rigid support
(Smid \& Novosad 1981; Jokati \& Moriyama 1979; Brockbank {\em et al.} 1997).
Such data shows unambiguously the
presence of a minimum (`the pressure dip') in the vertical normal stress
below the apex of the pile. With a plausible choice of secondary
c.r. (of which several were tried, with only minor differences
resulting), the {\sc FPA} case, in particular, was found to give a fairly good
quantitative account of the data of Smid \& Novosad (1981), and of
Brockbank {\em et al.} (1997); see Fig.~\ref{fig:dip}. This is
remarkable, in view of the radical simplicity of the assumptions made.

We accept, of course, that such models may be valid only a limited
regime in some larger parameter space. For example, since strain variables
are not introduced, these models cannot of themselves examine the
crossover to conventional elastic or elastoplastic behaviour that must
presumably arise when the applied stresses are significant on the scale
of the elastic modulus of the grains themselves. Some further remarks on
this crossover, in the context of anisotropic elastoplasticity, are made
in Section~\ref{sec:Strain}\ref{subsec:Anisotropic}.

In this paper we discuss the physical content of our general modelling
approach (of which the {\sc FPA} model is one example), based on
local stress propagation rules that depend on construction history, as
encoded in constitutive relations among stresses. In particular we
contrast the approach with more conventional ideas -- especially the
ideas of elastoplasticity.  For  simplicity, our mathematical discussion
is mainly limited to two dimensions (although our models were developed
to describe three-dimensional piles) and to the simplest, isotropic forms
of elastoplastic theory.  The discussion aims to sharpen some conceptual
issues. These concern not the details of particular models, but
the general question of what {\em sort} of description we should aspire
to: what sort of information do we need as modelling input, and what can
be predicted as output? An equally important (and closely related)
question is, what are the control variables in an experiment that must be
specified to ensure reproducible behaviour, and what are the observables
that can then be measured to depend on these?  For experiments on
sandpiles (briefly reviewed in Section~\ref{sec:Experiment})
we believe these to be open physics questions, and to challenge some
widely held assumptions of the applicability of traditional elastoplastic
modelling strategies to cohesionless poured grains.

The proposal that granular assemblies under gravity cannot properly be
described by the ideas of conventional elastoplasticity has been
opprobiously dismissed in some quarters: we stand accused of ignoring all
that is `long and widely known' among geotechnical engineers (Savage 1997).
However, we are not the first to put forward such a subversive proposal.
Indeed workers such as Trollope (1968) and Harr (1977) have long ago
developed ideas of force transfer rules among discrete
particles, not unrelated to our own approaches, which yield continuum
equations quite unlike those of elastoplasticity.  More recently, dynamical
{\em hypoplastic} continuum models have been developed (Kolymbas 1991,
Kolymbas and Wu 1993) which, as explained by Gudehus (1997) describe an
`anelastic behaviour without [the] elastic range, flow conditions and flow
rules of elastoplasticity'. Our own models, though not explicitly dynamic,
are similarly anelastic, as we discuss in Section~\ref{sec:Fragile}. They
should perhaps be classified as hypoplastic models, although their relation
to extremely {\em anisotropic} elastoplastic models is examined in
Section~\ref{sec:Strain}\ref{subsec:Anisotropic} below.

\section{Continuum models of cohesionless granular matter}
\label{sec:Continuum}
We start by reviewing (in their simplest forms) some well-known modelling
approaches based on rigid-plastic and elastoplastic ideas. This
is followed by a brief summary of the mathematical content of the
{\sc FPA} model and its relatives.

\subsection{Stress continuity and the Coulomb inequality}
The equations of stress
continuity express the fact that, in static equilibrium, the
forces acting on a small element of material must balance. For a
conical pile of sand we have, in
$d=3$  dimensions,
\be
\partial_r\srr + \partial_z\szr = \beta {(\scc-\srr)/
r}\label{1}\ee
\be\partial_r\srz + \partial_z\szz = g - \beta\srz/r\label{2}\ee
\be\partial_\chi\sigma_{ij} = 0\label{3}\ee where $\beta = 1$.
Here
$z,r$ and $\chi$ are cylindrical polar coordinates, with $z$ the
downward vertical. We take $r=0$ as a symmetry axis, so that
$\src=\szc=0$; $g$ is the force of gravity per unit volume;
$\sigma_{ij}$ is the usual stress tensor which is symmetric in
$i,j$. The equations for
$d=2$ are obtained by setting $\beta = 0$ in (\ref{1},\ref{2})
and suppressing (\ref{3}). These describe a wedge of constant
cross section and infinite extent in the third dimension. They also
describe a layer of grains in a thin, upright Hele-Shaw cell, but only if
the wall friction is negligible.

The Coulomb inequality states that, at any point in a cohesionless
granular medium, the shear force acting across any plane must be
smaller in magnitude than $\tan\phi$ times the compressive normal force.
Here $\phi$ is the angle of friction, a material parameter which,
in simple models, is equal to the angle of repose.
We accept this here, while noting that
(i) $\phi$ in principle depends on the texture (or fabric) of the
medium and hence on its construction history;
(ii) for a highly anisotropic packing, the existence of an
orientation-independent $\phi$ is questionable
(see Section~\ref{sec:Strain}\ref{subsec:Anisotropic});
(iii) the identification of $\phi$ with the repose angle ignores some
complications such as the
Bagnold hysteresis effect.

\subsection{Rigid-plastic models}
\label{subsec:rpm}
The model that Wittmer {\em et al.} ($1996,1997a$) refer to as ``incipient
failure  everywhere" ({\sc IFE}), is more commonly called the
rigid-plastic model. It postulates that the Coulomb condition is everywhere
obeyed {\em with equality} (Nedderman 1992).
That is, through every point in the material there passes
some plane across which the shear force is exactly $\tan\phi$
times the normal force.
By assuming this, the {\sc IFE} model allows closure
(modulo a sign ambiguity discussed below) of the
equations for the stress without invocation of an elastic strain field.
The {\sc IFE} model has therefore as its `constitutive relation'
(Wittmer {\em et al.} $1997a$):
\be
\sigma_{rr} = \sigma_{zz}\,{1\over \cos^2\phi}\left[
\sin^2\phi+1 \pm
2\sin\phi\,\,\sqrt{1-(\cot\phi\;\;\sigma_{zr}/\sigma_{zz})^2}\right]
\label{ife}
\ee
whereas the Coulomb inequality requires only that $\sigma_{rr}$
lies between the two values ($\pm$) on the right.

The postulate that a Coulombic slip plane passes through each and every
material point is not usually viewed as being accurate in
itself; the rigid-plastic model is more often proposed as a way of
generating certain `limit-state' solutions to an underlying elastoplastic
model. In the simplest geometries these solutions correspond to taking the
$-$ or $+$ sign in (\ref{ife}).
It is a simple exercise to show
that for a sandpile  at its repose angle, only one solution
of the resulting equations  exists in which the sign choice is everywhere
the same. This requires the negative root (conventionally referred to
as an `active' solution) and it shows a hump, not a dip, in the vertical
normal stress beneath the apex. Savage (1997), however, draws attention to a
`passive' solution, having a pronounced dip beneath the apex.
This solution actually contains a pair of matching
planes between an inner region where the positive root of (\ref{ife}) is
taken, and an outer region where the negative is chosen.

\begin{figure}[]
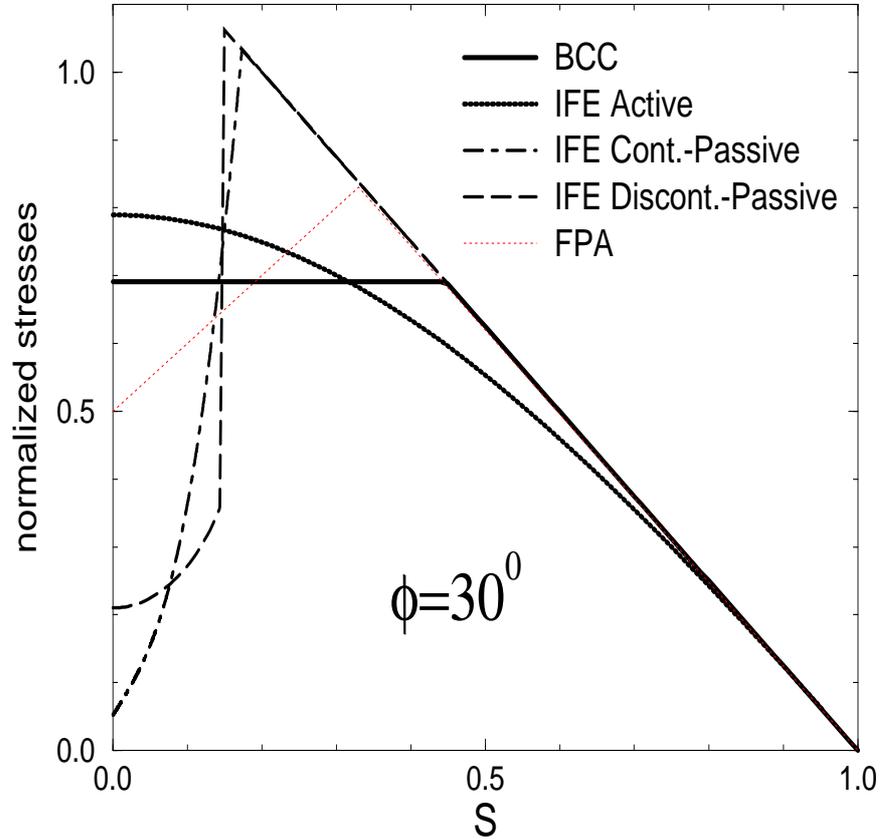

\caption{
Vertical normal stress found for the {\sc BCC} model
described in Section~\protect{\ref{sec:Continuum}\ref{bccmodel}}, for a pile
at angle of repose $\phi=30$ degrees, compared to the active and two passive
{\sc IFE} solutions (obtained by shooting from the midplane) discussed in
the text.  (Out of numerical reasons the continuous {\sc IFE} uses
$P=(\sigma_{zz}+\sigma_{xx})/2$ and the polar angle $\theta$ as functions
of the direction of the principal axis $\Psi$.)
Note that active and passive {\sc IFE} solutions do not bound the stress,
either in the {\sc BCC} model or in the simple elastoplastic model of
Cantelaube \& Goddard (1997), which, for a certain parameter choice,
yields identical results. 
The 2-dimensional {\sc FPA} solution is also included (dotted line).
\label{fig:bounds}}
\end{figure}

In principle there is more than one such `passive' solution. For example
one can seek an {\sc IFE} solution in which all stress components are
continuous across the matching plane. This requires a discontinuity in
the gradients of the stresses at the centreline
(see Fig.~\ref{fig:bounds}). The latter does not contradict Eq.~(\ref{ife}),
although it might be thought undesirable on other grounds (for example
if the {\sc IFE} equation is thought to bound the behaviour of a
simple elastoplastic body, for which the resulting displacement fields
might not be admissible). An alternative, which avoids this, is to
instead have a discontinuity of the normal stress parallel to the
matching plane itself.
This gives a second passive solution (Savage 1997,1998). These solutions do
not exhaust the repertoire of {\sc IFE} solutions  for the sandpile: there
is no physical principle that limits the number of matching surfaces. By
adding extra ones, a very wide variety of results can be achieved.

The emphasis placed on the rigid-plastic approach, at least in some
parts of the literature, seems to rest on
a misplaced belief that the limit-state solutions can be
`generally regarded as bounds between which
other states can exist, {\it i.e.}, when the material is behaving in an
elastic or elastoplastic manner' (Savage 1997).
A simple
counterexample is shown in Fig.~\ref{fig:bounds}. This shows the active and
two passive solutions (as defined above)
for a two-dimensional pile (wedge), along with an elementary elastoplastic
solution as presented recently by Cantelaube \& Goddard (1997)
and earlier by Samsioe (1955).
The latter is piecewise linear with
no singularity in the displacement field on the central axis; it happens
to coincide mathematically with the solution of a simple hyperbolic model
(Bouchaud {\em et al.} 1995) for the same geometry (there is no stress dip
in this particular model). Clearly the vertical normal stress does not
lie everywhere between that of the active and passive {\sc IFE} solutions,
which are therefore not bounds.

\subsection{Elastoplastic models}
\label{subsec:reasons}
In two dimensions at least, it has been argued that the pressure dip can be
explained within a simple conventional elastoplastic modelling approach.
This is certainly possible if the base is allowed to sag slightly
(Savage 1997). Here, however, we are concerned with piles built on a
rough, rigid support.
Even in this case, it has been argued that results similar
to those of the {\sc FPA} model can be obtained (Cantelaube \& Goddard 1997).
The simplest elastoplastic models assume a material
in which a perfectly elastic behaviour at small strains is conjoined
onto perfect plasticity (the Coulomb condition with equality) at larger
ones. In such an approach to the sandpile, an inner elastic region is
matched, effectively by hand, onto an outer plastic one. In the inner
elastic region the stresses obey the Navier equations,
which follow from those of Hookean elasticity by elimination of
the strain variables. The corresponding strain field is not
discussed, but tacitly treated as infinitesimal, since
the high modulus limit is taken. Howevever, for {\sc FPA}-like solutions,
which show a cusp in the vertical stress on the centreline, the
displacement shows singular features which are not easily reconciled with a
purely elastoplastic interpretation. The fact that the plastic
zone is introduced ad-hoc also has drawbacks -- for example it is hard to
explain the continued presence of such a zone if the angle of the pile is
reduced to slightly below the friction angle $\phi$
(to allow for the Bagnold hysteresis effect, say).
In {\sc OSL} approaches, an outer zone is not
assumed but predicted, and remains present in this case, although the
material in this zone is no longer at incipient failure.

The existence of {\sc FPA}-like
solutions to simple elastoplastic models in three dimensions, on a
non-sagging support, remains very much in doubt. But in any case,
our objections to the elastoplastic approach to modelling
sandpiles lie at a more fundamental level. Specifically it appears that,
to make unambiguous predictions for the stresses in a sandpile, these
models require boundary information that has no obvious physical meaning
or interpretation.  We return to this physics problem in
Section~\ref{sec:Strain}\ref{subsec:indet}.

\subsection{Local-rule models of stress propagation}
In the {\sc FPA} model (Wittmer {\em et al.} $1996, 1997a$)
one hypothesizes that, in each material
element, the orientation of the stress ellipsoid became `locked' into
the material texture at the time when it last came to rest, and does not
change in subsequent loadings (unless forced to: see
Section~\ref{sec:Fragile}).
This is a bold, simplifying assumption, and it may
indeed be far too simple, but it exemplifies the idea of having a local
rule for stress propagation that depends explicitly on construction
history.  For the sandpile geometry, where the material comes to rest on
a surface slip plane, this constitutive hypothesis leads to the
following relation  among stresses:
\be\srr = \szz -2\tan\phi \,\,\szr \label{5}\ee where $\phi$ is
the angle of repose. Eq.(\ref{5}) is algebraically specific to the case
of a pile created from a point source by a series of avalanches along
the free surface.

A consequence of Eq.~(\ref{5}), for a pile at repose, is that the
major principal axis everywhere bisects the angle between the
vertical and the free surface. It should be noted that in
cartesian coordinates, the {\sc FPA} model reads :
\be\sigma_{xx} = \szz -2\,\hbox{\rm sign}(x)\tan\phi
\,\,\sigma_{xz}
\label{4}\ee where $x$ is horizontal. From Eq.~(\ref{4}), the {\sc FPA}
constitutive relation is seen to be discontinuous on the central axis of
the pile: the local texture of the packing has a singularity on the
central axis which is reflected in the stress propagation rules of the
model. The paradoxical requirement, on the centreline, that the
principal axes are fixed simultaneously in two
different directions has a simple resolution:
the stress tensor turns out to be isotropic there.

The {\sc FPA} model is one of the larger class of {\sc OSL} models
in which the primary constitutive relation (in the sandpile
geometry) is, in Cartesians
\be \sigma_{xx} = \eta \szz + \mu\, \hbox{\rm sign}(x)\,\,\sigma_{xz}
\label{6}\ee
with $\eta,\mu$ constants.
As explained Wittmer {\em et al.} ($1997a$), these models (in two dimensions)
yield hyperbolic equations that have {\em fixed characteristic rays}
for stress propagation.
(Unless $\mu=0$, these are asymmetrically disposed about the vertical axis,
and invert discontinuously at the centreline
$x=0$.) The constitutive property that {\sc OSL} models describe
is that these characteristic rays, or force chains,  have orientations that
are `locked in' on burial of an element. The boundary condition, that
the free surface of a pile at its angle of repose $\phi$ is a slip plane,
yields one equation linking
$\eta$ and $\mu$ to  $\phi$; thus the {\sc OSL} scheme represents a
one-parameter family of models.
Note that, as soon as $\eta$ is not exactly unity (the {\sc FPA} case) the
orientation of the principal axes rotates smoothly as one passes through the
centreline of the pile. The assumption of fixed principal axes, though
appealing, is thus rather delicate, and arguably much less important than
the idea of fixed characteristics, since these represent the average
geometry of force chains in the medium.
The experimental data
(Fig.~\ref{fig:dip}) supports models in the {\sc OSL} family with
$\eta$ close, but perhaps not exactly equal, to unity.

\label{bccmodel}
Note that, unless the {\sc OSL} parameter is chosen
so that $\mu = 0$, a constitutive singularity on the central axis
remains. The case $\mu =0$ corresponds to one studied
earlier by Bouchaud {\em et al.} (1995);
this `{\sc BCC}' model is the only member of the {\sc OSL} family to have
characteristics symmetric about the vertical.
(Their angles $\pm\theta$ to the vertical obey $\tan^2\theta = c_0^2 = \eta$.)
This latter model could be called a `local Janssen model' in that it assumes
{\em local} proportionality of horizontal and vertical compressive stresses
-- an assumption which, when applied {\em globally} to average stresses in a
silo, was first made by Janssen (1895).

The local-rule models just discussed do not account for the presence of
`noise' or randomness in the local texture. Such effects have been
studied by Claudin {\em et al.} (1998), and, if the noise level is not
too large, lead at large length scales to effective wavelike equations with
additional gradient terms giving a diffusive spreading of the
characteristic rays.
The limit where the diffusive term dominates
corresponds to a {\em parabolic} differential equation (Harr 1977),
similar to those arising in scalar force models (Liu {\em et al.} 1995)
which have, in effect, a single downward
characteristic (so the main interest lies in the diffusive spreading).
It is possible that under extreme noise levels this picture changes again,
although this conclusion is based on assumptions about the noise itself
that may not be valid (Claudin {\em et al.} 1998).
The discussions that follow therefore apply to local-rule models with
moderate, but perhaps not extreme, noise.

Note finally that the fact that two continuum models, based on different
constitutive hypotheses, can give identical results for the stresses in some
specified geometry, obviously does not mean that the models are equivalent.
(Equivalence requires, at least, that the Green function of the two models
is also the same.) Thus, for example, models such as {\sc FPA}
are not equivalent to Trollope's model of ``clastic discontinua"
(Trollope 1968, Trollope and Burman 1980). In Appendix \ref{sec:Trollope}
we outline the relationship between our work and the marginal packing
models studied by Ball \& Edwards (to be published), Huntley (1993), Hong
(1993), Bagster (1978) and Liffmann {\em et al.} (1992), as well as
the work of Trollope (1956,1968).

\section{Strain and displacement variables}
\label{sec:Strain}

In the {\sc FPA} model and its relatives, strain variables are not considered.
A partial justification for this was given by
Wittmer {\em et al.} ($1997a$),
namely that the experimental data obey a form of radial stress-field
({\sc RSF}) scaling: the stress patterns observed at the base are the same
shape regardless of the overall height of the pile. Formally one has
for the stresses at the base
\be
\sigma_{ij} = gh\, s_{ij}(r/ch) \label{rsf}
\ee
where $h$ is the pile height, $c = \cot\alpha$  and $s_{ij}$ a reduced
stress: $\alpha$ is the angle between the free surface and the horizontal so
that for a pile at repose,
$\alpha = \phi$.
This form of {\sc RSF} scaling, which involves only the forces at the base
(Evesque 1997), might be called the `weak form' and is
distinct from the `strong form' in which Eq.~(\ref{rsf}) is obeyed
also with $z$ (an arbitrary height from the apex) replacing $h$ (the
overall height of the pile).

This scaling implies that there is no
characteristic length-scale. Since elastic deformation introduces
such a length-scale (the length-scale over which an elastic pile
would sag under gravity) the observation of {\sc RSF} scaling to
experimental accuracy suggests that elastic effects {\em
need not be considered explicitly}.
We accept however (correcting Wittmer {\em et al.} $1997a$)
that this does not of itself rule out
elastic or elastoplastic behaviour which, at least in the limit of
large modulus, can also yield equations for the stress from which the bulk
strain fields cancel. Note that it is tempting, but entirely wrong, to
assume that a similar cancellation occurs at the boundaries of the
material; we return to this below (Section \ref{sec:Boundary}).

The cancellation of bulk strain fields in elastoplastic models is
convenient since there appears to be no clear definition of strain or
displacement for piles constructed by pouring sand grainwise from a
point source. To define a physical displacement or strain field, one
requires a reference state.
In (say) a triaxial strain test (see e.g. Wood 1990) an initial state 
is made by some reproducible procedure, and strains measured from there.
The elastic part is identifiable in principle, by removing the applied
stresses (maintaining an isotropic pressure) and seeing how much the
sample recovers its shape.
In contrast, a pile constructed by pouring grains onto its apex
cannot convincingly be described in terms of the plastic and/or
elastic deformation from some initial reference state of the same
continuous medium: the corresponding experiments are
unrealizable. Even were the load
(which consists purely of gravity) to be removed, the resulting
unstrained body would comprise grains floating freely in space with no
definite positions. It is unsatisfactory to define a strain or
displacement field with respect to such a body. The problem occurs whenever
the solidity of the body itself {\em only} arises because of the
load applied. A similar situation occurs, for example, in colloidal
suspensions that flow freely at small shear stresses but (by jamming) can
support larger ones indefinitely (Cates {\em
et al.}, to be published).

Although one cannot uniquely define the strain in a granular assembly
under gravity, it may of course be possible to define {\em incremental}
strains in terms of the displacement of grains when a small load is added.
However, the range of stress increments involved might in practice be
negligible (Kolymbas 1991; Gudehus 1997).

\section{The role of boundary conditions}
\label{sec:Boundary}

\subsection{Boundary conditions in hyperbolic (and parabolic) models}
\label{subsec:charz}
Models that assume local
constitutive equations among stresses (including all {\sc OSL} models,
and also the {\sc IFE} or rigid-plastic model) provide hyperbolic differential
equations for the stress field. Accordingly, if one specifies a
zero-force boundary condition at the free (upper) surface of a
wedge, then any perturbation arising from a small extra body
force (a `source term' in the equations) propagates along two
characteristics passing through this point. (In the {\sc OSL} models
these characteristics are, moreover, straight lines.)
Therefore the force at the base can be found simply by
summing contributions from all the body forces as propagated along
two characteristic rays onto the support; the sandpile problem is, within
the modelling approach by
Bouchaud {\em et al.} (1995) and Wittmer {\em et al.} ($1996,1997a$),
mathematically well-posed.

\begin{figure}[]
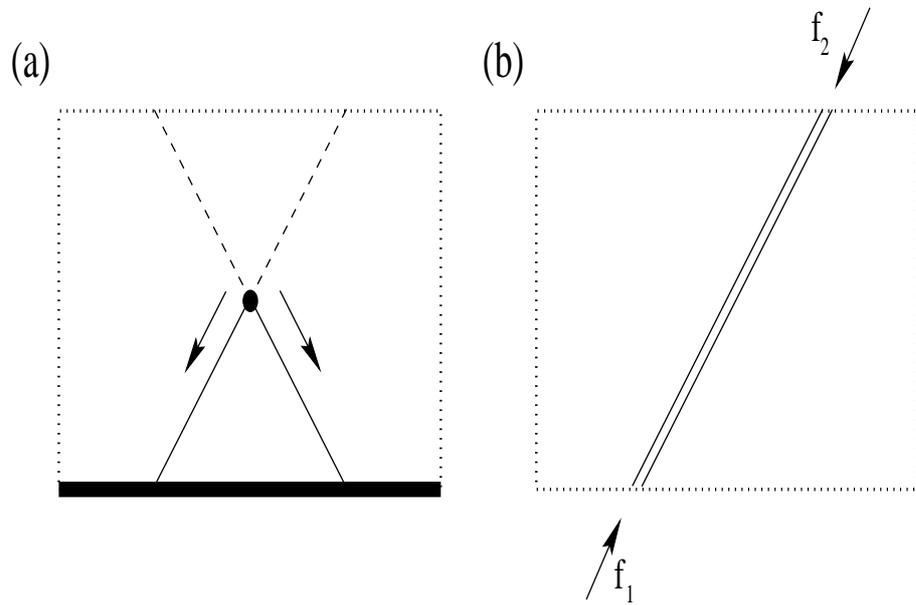

\caption{(a) The response to a localized force is found by resolving it along
characteristics through the point of application, propagating along those
which do
not cut a surface on which the relevant force component is specified.
For a pile under gravity, propagation is only along the downward rays.
(b) Admissible boundary conditions cannot specify separately the force
component at both ends of the same characteristic. If these forces
are unbalanced (after allowing for body forces), static equilibrium is lost.
\label{fig:pathfig}}
\end{figure}

Note that in
principle, one could have propagation also along the `backward'
characteristics (see Fig.~\ref{fig:pathfig}(a)). This is forbidden since
these cut the free surface; any such propagation can only arise in the
presence of a nonzero surface force, in violation of the boundary
conditions. Therefore the fact that the propagation occurs only along
downward characteristics is not related to the fact that gravity acts
downward; it arises because we know already the forces acting at the free
surface (they are zero). Suppose we had instead an inverse problem:
a pile on a bed with some unspecified overload at the top surface, for
which the forces acting at the base had been measured. In this case, the
information from the known forces could be propagated along the {\em
upward} characteristics to find the unknown overload.
More generally, in {\sc OSL} models of the sandpile, each characteristic ray
will cut the surface of a (convex) patch of material at two points. Within
these models, the sum of the forces tangential to the ray at the two ends
must be balanced by the tangential component of the body force integrated
along the ray (see Fig.~\ref{fig:pathfig}(b)).
We discuss this physics (that of force chains)
in Section~\ref{sec:Fragile}\ref{subsec:Stresspaths}.

In three dimensions, the mathematical structure of these models is
somewhat altered (Bouchaud {\em et al.} 1995), but the
conclusions are basically unaffected. Note however that for different
geometries, such as sand in a bin, the problem is not well-posed even with
hyperbolic equations, unless something is known about the interaction
between the medium and the sidewalls.
Ideally one would like an approach in which sidewalls and base were
treated on an equal basis;  this is the subject of ongoing research.
Note also that the essential character of the boundary value problem
is not altered when appropriate forms of randomness are introduced.
For although the response to a point force is now spread about the two
characteristics, even in the parabolic limit (where the underlying
straight rays are effectively coincident and only spreading remains)
the sandpile boundary value problem remains well posed.

\subsection{The physics of elastic indeterminacy}
\label{subsec:indet}
The well-posedness of the sandpile does not extend
to  models involving the elliptic equations for an elastic body.
For such a material, the stresses throughout the body can be solved only
if, at all points on the boundary, either the force distribution
or a displacement field is specified
(Landau \& Lifshitz 1986).
Accordingly, once the zero-stress boundary condition is applied at the
free surface, nothing
can in principle be calculated unless either the forces or
the displacements at the base are already known (and the former amounts
to specifying in advance the solution of the problem). 
From an elastoplastic perspective, it is clearly absurd to try to calculate
the forces on the support, which are the experimental data, without
some further information about what is happening at the bottom boundary.
We have called this the problem of
`elastic indeterminacy' (Bouchaud {\em et al.} 1998)
although perhaps `elastic ill-posedness' would be a better term.
The problem does not arise from any uncertainty about what to do
mathematically: one should specify a
displacement field at the base. Difficulties nonetheless arise if, as we
argued above, no physical significance can be attributed to this
displacement field for cohesionless poured sand.

\begin{figure}[]
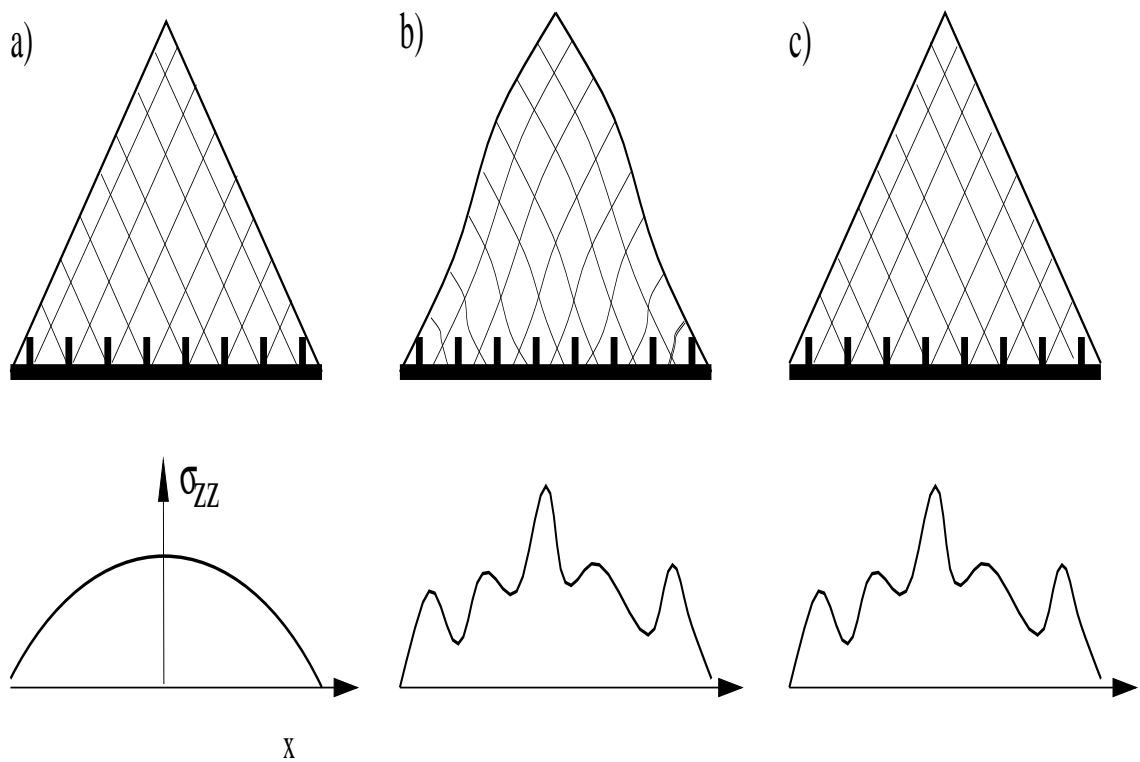

\caption{
Starting from an elastic cone or wedge on a rough
support, any initial stress distribution can be converted to
another by displacements with respect to the rough `pinning'
surface (a) $\to$ (b). Taking the limit of a high modulus (b)
$\to$ (c) at fixed surface forces, an arbitrary stress field remains,
while recovering the initial shape of the cone and satisfying the free
surface boundary conditions. This shows the physical character of
`elastic indeterminacy' for an elastic or elastoplastic body on
a rough support.
\label{fig:indeterminacy}}
\end{figure}

To give a
specific example, consider the static equilibrium of an elastic
cone or wedge of finite modulus resting on a completely rough, rigid
surface (which one could visualize as a set of pins;
Fig.~\ref{fig:indeterminacy}).
Starting from any initial configuration, another can be generated
by pulling and pushing parts of the body horizontally across the
base ({\em i.e.}, changing the displacements there); if this is rough,
the new state will still be pinned and will achieve a new static
equilibrium. This will generate a stress distribution, across the
supporting surface and within the pile, that differs from the
original one. If a large enough modulus is now taken
(at fixed forces), this procedure allows one to generate arbitrary
differences in the stress distribution while generating
neither appreciable distortions in the shape of the cone, nor any
forces at its free surface. Analogous remarks apply to any simple
elastoplastic theory of sandpiles, in which an elastic zone, in
contact with part of the base, is attached at matching surfaces
to a plastic zone.

In contrast, experimental reports
(reviewed in Section~\ref{sec:Experiment}) indicate that
for sandpiles on a rough rigid support, the forces on the base can be
measured reproducibly. They also suggest that these forces, although
subject to statistical fluctuations on the scale of several grains,
do not vary too much from one pile to another, at
least among piles constructed in the same way
(e.g., by avalanches from a point source), from the same material.
This argues strongly against the idea that such forces in fact depend on a
basal displacement field, which is determined either by the whim of the
experimentalist, or by some as-yet unspecified physical mechanism acting
at the base of the pile. Note that basal sag
is {\em not} a candidate for the missing mechanism, since
it does not resolve the elastic indeterminacy in these models; the
latter arises primarily from the {\em roughness}, rather than the
rigidity, of the support.
Note also, however, that elastic indeterminacy can be alleviated in
practice if the elastoplastic model is sufficiently anisotropic; we return
to this point in Section~\ref{sec:Strain}\ref{subsec:Anisotropic}.

Evesque (private communication), unlike many authors, does confront
the issue of elastic indeterminacy and seemingly concludes that
the experimental results {\em are and must be indeterminate}; he
argues that the external forces acting on the base of a pile can
indeed be varied at will by the experimentalist, without causing
irreversible rearrangements of the grains
(see also Evesque \& Boufellouh 1997).
To what extent this viewpoint is based on experiment, and to what extent
on an implicit presumption in favour of elastoplastic theory, is to us
unclear.

\subsection{Displacements to the rescue?}
\label{subsec:postul}
Let us boldly suppose, then, that the experimental data is meaningful and
reproducible, at least as far as the global,
`coarse-grained' features of the observations are concerned.
(Noise effects at the level of individual grains
may in contrast be exquisitely sensitive to temperature and other
poorly-controlled parameters; Claudin \& Bouchaud 1997.)
Adherents to traditional elastoplastic models then have three choices.
The first is to consider the possibility that,
after all, the problem of cohesionless poured
sand may be better described by quite different
governing equations from those of simple elastoplasticity. This possibility,
which represents our own view, has certainly been
suggested before. 
For example, hypoplastic models in which there is negligible elastic range 
(Gudehus 1997; Kolymbas 1991, Kolymbas and Wu 1993) do not suffer from 
elastic indeterminacy.

The second choice is to postulate various additional constraints, so as to
eliminate some of the infinite variety of solutions that elastoplastic
models allow (unless basal displacements are specified). For
example, it is tempting to impose (in its strong form) {\sc RSF} scaling:
for a wedge, as shown by Samsioe (1955) and Cantelaube \& Goddard (1997) this
reduces the admissible solutions to a piecewise linear form.
Such a postulate may seem quite harmless: after all, we have
emphasized already that the observations do themselves show (weak) {\sc RSF}
scaling. However, according to these models, the
central part of the pile can correctly be viewed
as an elastic continuum; hence from any solution
for the stresses it {\em should be} physically
possible to generate another by an infinitesimal
pushing and pulling of the elastic material along
the rough base. Accordingly one has no reason to expect even the weak
{\sc RSF} scaling observed experimentally. Setting this aside, one could
impose weak {\sc RSF} scaling by assuming a basal displacement
field of the same overall shape for piles of all sizes. However, as pointed
out by Evesque (1997), even this imposition does not require the
{\em strong} form of {\sc RSF} scaling assumed by Cantelaube
\& Goddard (1997).  In summary, simple elastoplastic models of sandpiles {\em
require} that the experimental results for the force at the base depend on
how the material was previously manipulated. Any attempt to
predict the forces without specifying these manipulations is misguided.

\begin{figure}[]
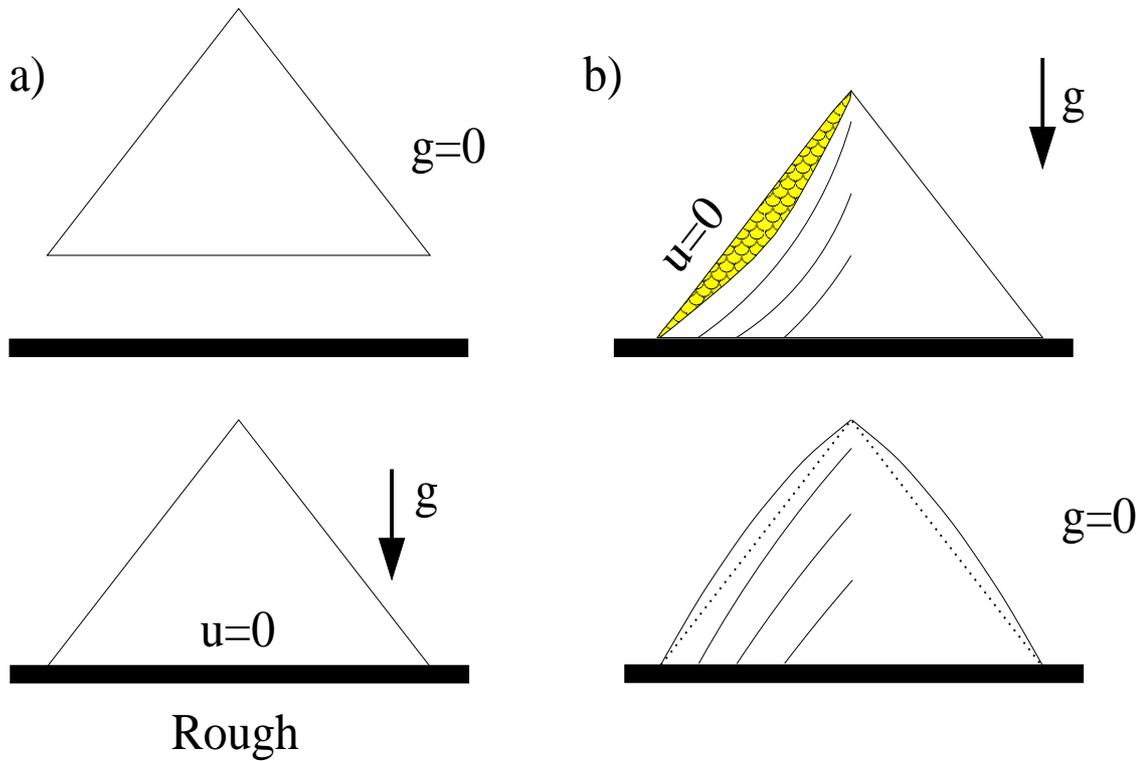

\caption{
(a) The `spaceship' model of a sandpile.
An unstrained, isotropic elastoplastic cone or
wedge is brought into contact with a rough surface and gravity then
switched on.
(b) `Laminated elastoplastic model' of a sandpile. Layers
are added in a state of zero stress (thereby, it is argued, zero
displacement) to a pre-existing, gravitationally loaded pile.
Such a pile (if gravity is removed) will spring into a new shape,
characterized by a nonzero internal stress field 
(Appendix~\ref{sec:Trollope}).
\label{fig:spaceship}}
\end{figure}

A third reaction, therefore, is to start modelling explicitly the physical
processes going on at the base of the pile. As mentioned previously, one is
required to specify a displacement field at the base of the
elastic zone; more accurately, it is the product of the displacement field
and the elastic modulus that matters. This need not vanish in the large
modulus limit (Section~\ref{sec:Strain}\ref{subsec:indet});
one possible choice, nonetheless, is
to set the displacement field to zero at a finite modulus (which might then
be taken to infinity). The simplest interpretation of this choice is by
appeal to a model in which the `sandpile' is constructed as follows
(Fig.~\ref{fig:spaceship}(a)): an elastoplastic wedge, floating freely in
space, is brought to rest in contact with a rough surface, in a state of
zero strain.
Once in contact, gravity is
switched on with no further adjustments in the contact region allowed.
This might be referred to as the `spaceship model' (or perhaps the
`floating model') of a sandpile. This illustrates two facts: (a) in
considering explicitly the displacement field at the bottom surface,
elastoplastic modellers are obliged to make definite
assumptions about the previous history of the material;  (b) these
assumptions do not usually have much in common with the actual
construction history of a sandpile made by pouring.
A possible alternative to the
spaceship model, in which unstressed laminae of elastoplastic material are
successively added to an existing pile (Fig.~\ref{fig:spaceship}(b)) is
discussed in Appendix~\ref{sec:Laminated}.

\subsection{Determinacy and anisotropy}
\label{subsec:Anisotropic}
We shall now show that hyperbolic behaviour can be recovered
from an elastoplastic description by taking a
strongly anisotropic limit (Cates {\em et al.}, to be published).
For simplicity we restrict attention to the {\sc FPA} model.

The {\sc FPA} model describes, by definition, a material in which the shear
stress must vanish across a pair of orthogonal planes fixed in the
medium -- those normal to the (fixed)
principal axes of the stress tensor. According to the Coulomb
inequality, which the model also adopts, the shear stress must also be
less than
$\tan\phi$ times the normal stress, across planes oriented in all other
directions.  Clearly this combination of requirements can be viewed as a
limiting case of an elastoplastic model with an anisotropic yield
condition:
\be
|\sigma_{tn}| \le \sigma_{nn}\,\tan\Phi(\theta) \label{yieldcrit}
\ee
where $\theta$ is the angle between the plane normal ${\bf n}$ and the
vertical (say).
An anisotropic yield condition should arise, in principle, in any
material having a nontrivial fabric, arising from its construction
history. The limiting choice corresponding to the {\sc FPA} model for a
sandpile is
$\Phi(\theta) = 0$  for $\theta = (\pi - 2\phi)/4$ (this corresponds to
planes where $\bf n$ lies parallel to the major principal axis), and
$\Phi(\theta) = \phi$ otherwise. 
(There is no separate need to specify the second,
orthogonal plane across which shear stresses vanish, since this is
assured by the symmetry of the stress tensor.)
By a similar argument, all other {\sc OSL} models can also be
cast in terms of an anisotropic yield condition, of the form $|\sigma_{tn} -
\sigma_{nn}\,\tan\Psi(\theta)| \le \sigma_{nn}\tan\Phi(\theta)$ where
$\Phi(\theta)$ vanishes, 
and $\Phi(\theta)$ is finite for two values of $\theta$.
(This fixes a {\em nonzero} ratio of shear and normal
stresses across certain special planes.)

At this purely phenomenological level there is no difficulty in connecting
hyperbolic models smoothly onto {\em highly anisotropic} elastoplastic
descriptions.
Specifically, consider a medium having an orientation-dependent
friction angle $\Phi(\theta)$ that does not actually vanish,
but is instead very small
($\le \epsilon$, say) in a narrow range of angles (say of order
$\epsilon$) around $\theta = (\pi - 2\phi)/4$, and approaches $\phi$
elsewhere.  (One interesting way to achieve the required yield anisotropy is
to have a strong anisotropy in the {\em elastic} response, and then impose a
{\em uniform} yield condition to the strains, rather than stresses.)

Such a material will have, in principle, mixed
elliptic/hyperbolic equations of the usual elastoplastic type. The
resulting elastic and plastic regions must nonetheless arrange
themselves so as to obey the {\sc FPA} model to within terms that vanish as
$\epsilon \to 0$. If
$\epsilon$ is small but finite, then for this elastoplastic
model the results  will depend on the basal boundary condition,
but only through these higher order corrections to the leading ({\sc FPA})
result. We show in Section~\ref{sec:Fragile} below that the case of small but
finite
$\epsilon$ is exactly what one would expect if a small amount of
particle deformability were introduced to a fragile skeleton of rigid
particles obeying the {\sc FPA} constitutive relation.

Although somewhat contrived (from an elastoplastic standpoint), the above
choice of anisotropic yield condition establishes an
important point of principle, and may point toward some important new
physics. Although elastoplastic models do suffer
from elastic indeterminacy (they require a basal displacement field to be
specified), the extent of the influence of the boundary condition on the
solution depends on the model chosen. Strong enough
(fabric-dependent) anisotropy, in an elastoplastic description, might so
constrain the solution that, although it suffers elastic
indeterminacy in principle, it does so only harmlessly in practice.
Under such conditions it is {\em primarily the fabric} and only
minimally the boundary conditions which actually determine the stresses
in the body. For models such as that given above there is a
well-defined limit where the indeterminacy is entirely lifted,
hyperbolic equations are recovered, and it is quite proper to talk of
local stress propagation `rules' which are determined, independently of
boundary conditions, by the fabric (hence
construction history) of the material.

Our modelling framework, based precisely on these assumptions, will
be valid for sandpiles if, as we contend, their physics lie close to this
limit of `fabric dominance' (see Section~\ref{sec:Fragile} below).
This contention is consistent with, though it does not
require, belief in the existence of an underlying 
elastoplastic continuum description.

\section{Experimental results}
\label{sec:Experiment}
Before discussing
in more detail the physical interpretation of our models, we
give a brief account of the experimental data. In doing this, it
is important to draw a distinction between (axially symmetric) cones,
and (translationally symmetric) wedges of sand. The latter is a
quasi-two dimensional geometry.
The main question is, to what extent the pressure-dip can be trusted as a
reproducible experimental phenomenon for a sandpile constructed by pouring
onto a rough rigid support.
In particular, Savage (1997) has drawn attention to
the possible role of small deflections in the base (`basal sag') in
causing the dip to arise.

\subsection{Cones}
\begin{figure}[]
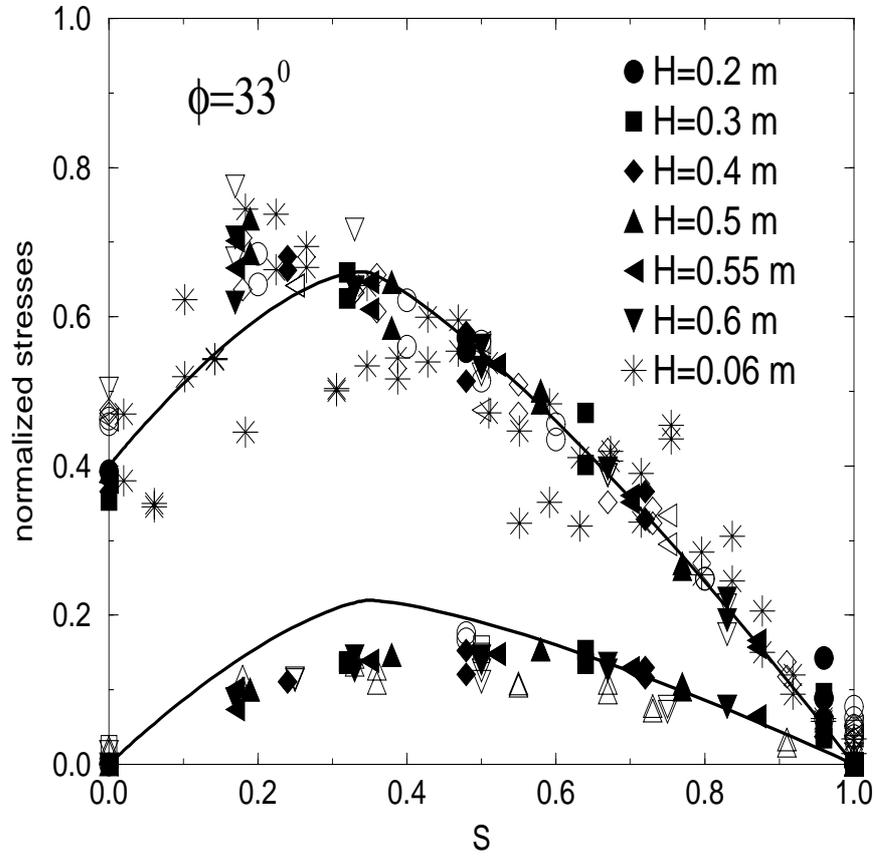

\caption{Comparison of {\sc FPA} model
using a uniaxial secondary closure (Wittmer {\em et al.} 1996, $1997a$)
with scaled experimental data of Smid \& Novosad (1981) and
\mbox{(*)} that of Brockbank {\em et al.} (1997)
which was averaged over three piles.
Upper and lower curves denote normal and shear stresses.
The data is used to calculate the total weight of the pile which is then
used as a scale factor for stresses. The horizontal coordinate is scaled
by the pile radius.
\label{fig:dip}}
\end{figure}

The earliest data we know of, on conical sandpiles, 
is that of Hummel \& Finnan (1920) who observed
a pronounced stress dip. However, their pressure cells were apparently
subject to extreme hysteresis, and these results cannot be relied upon.
Otherwise the only data prior to Smid \& Novosad (1981) {\em for cones} is
that of Jokati \& Moriyami (1979). Although a stress dip is
repeatedly observed by these authors, their results (on rather small piles)
do not show consistent {\sc RSF} scaling.
The well-known data of Smid \& Novosad (1981) shows a clear stress minimum
at the centre of the pile. Even this dataset is not completely satisfactory:
the observation of the dip is based on the data from a single (but calibrated)
pressure cell beneath the apex. However, the data for different pile heights
shows clear (weak) {\sc RSF} scaling, and is quantitatively fit by the
{\sc FPA}
model with either of the secondary closures shown in Fig.~\ref{fig:dip}.
Savage (1997) points out that
`it is not possible from the information given to estimate the deflections
[at the base] that might result from the weight of the pile'.
Smid and Novosad, however, describe their platform as `rigid'.
\footnote{
Savage (1997) also criticises the reduction method used to analyse 
this data by Wittmer {\em et al.} ($1997a$), as shown in Fig.~\ref{fig:dip}; 
when normalizing stresses by the mean density of the pile, he apparently
prefers to use a separate measurement of the bulk density (in a different
geometry), rather than the density deduced by integrating the vertical normal
stresses to give the weight of the pile.
}

Recently, Brockbank {\em et al.} (1997) have performed
a number of careful measurements on relatively small piles of sand
(as well as flour, glass beads, etc.).
The pressure transducers comprise an assembly of steel
ball-bearings lying atop a thin blanket of transparent rubber on
a rigid glass plate; material is poured from a point source onto this
assembly. The deflection of the ball-bearings is estimated as 10
$\mu$m. By calibrating and optically monitoring their imprints on
the rubber film, the vertical stresses can be measured.
Perhaps the most interesting feature of this method is that,
although the basal deflection is certainly not zero, it is of a
character quite unlike basal sag. Indeed, the supporting
ball-bearings are deflected downward (indenting the rubber film)
in a manner that depends on the {\em local} compressive stress,
as opposed to the cumulative ({\em i.e.}, nonlocal) effect of
sagging. The latter is bound to be maximal under the apex of the
pile, whereas the indentation is maximal under the zone of
maximum vertical compressive stress, wherever that may be.
If the stress pattern is controlled by slight
deformations of the base, there would be no reason to expect a
similar stress pattern to arise for an indentable base, as for
a sagging one.

But in fact, a very similar stress pattern is seen (Fig.~\ref{fig:dip}).
The data shown here involve averaging over several piles, since the
setup measures stresses over quite small areas of the base (the
ball bearings are 2.5 mm diameter) and these stresses fluctuate locally, as
is well-known (Liu {\em et al.} 1995; Claudin \& Bouchaud 1997).
Although still subject to relatively large statistical scatter,
the data show an unambiguous dip of very similar magnitude to that
reported by Smid and Novosad; moreover the dip is spread over
several, rather than a single, transducer(s).

It is, of course, important to distinguish conceptually the
noisiness of this data (arising from fluctuations
at the granular level) from any intrinsic irreproducibility of the results.
If the results are reproducible, then for large enough piles one
might expect the averaging over several piles to be
obviated by binning the data over many transducers.
This is, in effect, what Smid and
Novosad do (since their transducers are much
larger). More careful experimental investigations
of this point would, nonetheless, be welcome.

We conclude from this recent study, which substantially
confirms the earlier work of Smid \& Novosad (1981),
that the attribution of the stress dip to basal sag
is not justified
for the case of conical piles of sand.
Brockbank {\em et al.} (1997) also saw a stress dip
for small, but not large, glass beads. This difference
suggests that to observe the dip requires a large enough pile compared to
the grain size -- perhaps to allow an anisotropic mesoscale texture to
become properly established.
No dip was seen for lead shot (deformable) or flour (cohesive).

\subsection{Wedges}
\label{subsec:Wedges}
The experiments on {\em wedges} appear very different. The
papers of Hummel \& Finnan (1920), and Lee \& Herington (1971) include
datasets for which the construction history is described as being
effectively from a line source.  These results, as well as others cited by
Savage (1997) offer support for his conclusions (made earlier by Burman
\& Trollope 1980) that the construction history of the wedge does not much
matter,  and that there is only a very small or negligible dip for wedges
supported  by a fully rigid base. These studies also suggest that a dip
appears almost immediately if the base under the wedge is allowed to sag.
These results, if confirmed by careful repetition of the experiments,
would certainly cast doubt on {\sc FPA}-type models as applied to wedges.

\subsection{Specifying the construction history}
\label{subsec:plasticone}
Such historic experiment, measuring the
stress distribution for wedges made supposedly from a line source, need
careful repetition. This is
because, even from a point source (conical pile) or line source (wedge)
at least two different types of construction history are possible.
The first is when, as assumed in {\sc FPA}-type models, the grains avalanche
in a thin layer down the free surface.
The second, which, like the first, has clearly been observed in
three-dimensional work on silo filling
(Munch-Andersen \& Nielsen 1990, 1993)
is called `plastic cone' behaviour.
It entails the impacting grains forcing their way downwards at the apex
into the body of the pile, which then spreads sideways. A
parcel of grains arriving at the apex ends up finally as a thin horizontal
layer.
(A transition between this and surface avalanche flow may be
controlled by varying the height from which grains are dropped, among
other factors.)
A third possibility is that of `deep yield' (see Evesque \& Boufellouh 1997):  
a buildup of material near the apex followed by a deep avalanche in
which  a thick slab of material slumps outwards (Evesque 1991).

These different construction histories, even among piles created from
a point or line source, would lead one to expect quite different stress
patterns. For example, the plastic cone construction should lead to a
texture with local symmetry about the vertical, as assumed by
Bouchaud {\em et al.} (1995).
This model, which we also expect to describe a conical
pile built by sieving sand uniformly onto a supporting disc
(Wittmer {\em et al.} $1997b$) does not give a pressure dip.
Although in point-source experiments on cones the surface avalanche
mechanism is usually seen (Evesque 1991; Evesque {\et al.} 1993)
we do not know whether the same applies for wedges; the classical
literature is ambiguous (Hummel \& Finnan 1920; Lee \& Herington 1971).
For these reasons such experiments must be repeated, with proper monitoring
of the construction history, before conclusions can be drawn.

There are, in fact, good reasons why the surface avalanche scenario, on
which models such as {\sc FPA} depend, may be very hard to observe in the
wedge geometry.
Recall that for the wedge geometry at repose, all {\sc OSL} models predict
an outer sector of the wedge, of substantial thickness,
in which the Coulomb inequality is saturated.
Clearly, if avalanches take place on top of a thick slab of
material already at incipient failure, it may be impossible to avoid
rearrangements deeper within the pile, leading either to `deep yield' or
`plastic wedge' behaviour.
To this extent the application of {\sc FPA}-type models to a wedge geometry
is not necessarily self-consistent.
The same does not apply in the
conical geometry, where the solution of these models predicts only an
infinitesimal plastic layer at the surface of the cone
(Wittmer {\em et al.} 1996).
Accordingly
it would be very interesting to compare experimentally wedges and cones of
the same material to see whether the character of the avalanches is
fundamentally different, as {\sc FPA}-like models might lead one to expect.
Further experiments involving comparison of histories are suggested by
Wittmer {\em et al.} ($1997a,b$).

Although there are, so far, few data showing a clear dependence
of measured stresses on construction history in freestanding cones or wedges,
the effect is well-established in experiments on silos. Specifically, for
flat-bottomed silos filled by surface avalanches from a point source, the
vertical normal force at the centre of the base is less than at the edge
(Munch-Andersen \& Nielsen 1990).  This effect, which is readily explained
within an {\sc FPA}-type modelling approach (Wittmer {\em et al.} $1997a$),
is not reported in silos filled
by sieving, nor when a plastic cone behaviour is seen at the
apex (Munch-Andersen \& Nielsen 1993).

\section{Sandpiles as fragile matter}
\label{sec:Fragile}
As we have emphasized, the continuum
mechanics represented by our hyperbolic models is not that of
conventional elastoplasticity.
In what follows we develop an outline interpretation of this continuum
mechanics as that appropriate to a material in which stresses propagate
primarily along force chains.
Simulations of frictional spheres offer some support for the
force-chain  picture, at least as a reasonable approximation: most of the
deviatoric stress is found to arise from {\em strong, normal} forces between
particles participating in force chains; tangential forces (friction) and
the weaker contacts transverse to the chains contribute mainly to the
isotropic pressure
(Thornton \& Sun 1994; Thornton 1997;
C.~Thornton, this volume).  In addition to this,
the content of our models is to assume that the
skeleton of force chains is {\em fragile}, in a
specific sense defined below.

\subsection{Force chains}
\label{subsec:Stresspaths}

Informally speaking, the hyperbolic problem posed by {\sc OSL} models is
determined once half of the boundary forces are specified. More precisely
(Fig.~\ref{fig:pathfig}(b)) one is required to specify the surface force
tangential to each characteristic ray, at one end and {\it one end only}. The
corresponding force acting at the other end is obliged to balance the sum
of the specified force, any body forces acting tangentially along the ray. If
it does not do so, then within our modelling approach, the material ceases
to be in static equilibrium. This is no different from the corresponding
statement for a fluid or liquid crystal; if boundary conditions are
applied that violate the conditions for static equilibrium, some sort of
motion results. Unlike a fluid, however, for a granular medium we  expect
such motion to be in the form of a finite rearrangement rather than a
steady flow. Such a rearrangement will change the microtexture of the
material, and thereby {\it alter the constitutive relation among
stresses}. We expect it to do so in such a way that the new network of
force chains (new constitutive relation) is able to support the newly
imposed forces.

\begin{figure}[]
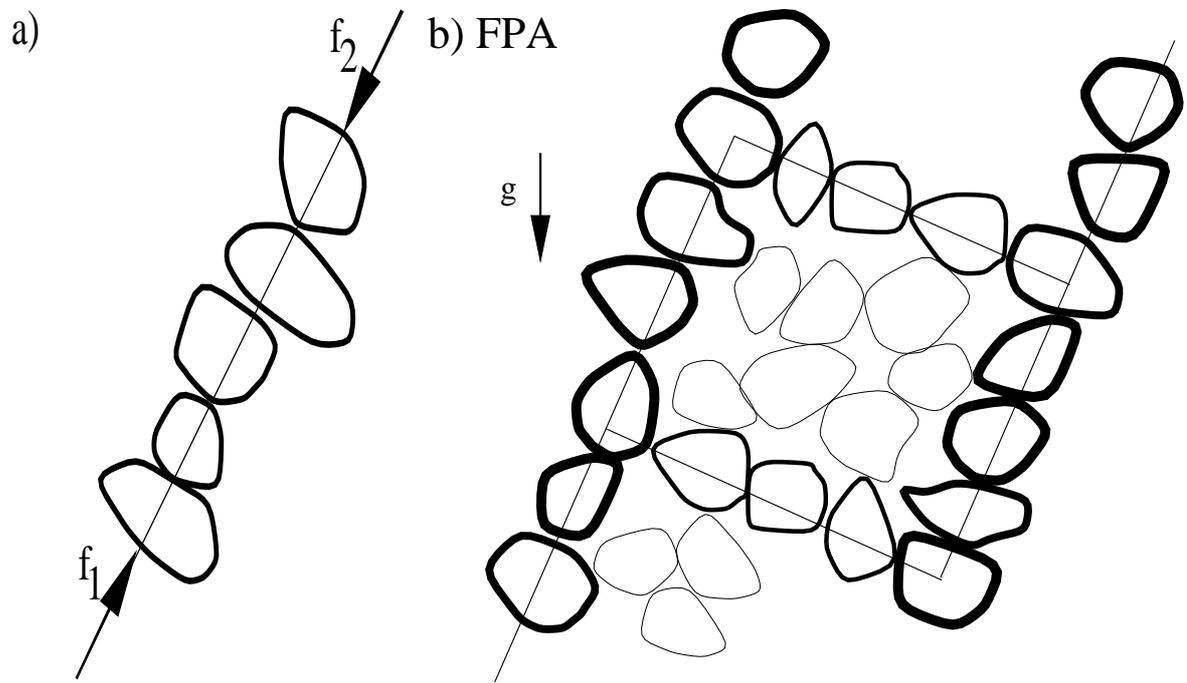

\caption{
(a) A force chain (``stress path") of hard particles can support
only tangential compressive loads in static equilibrium. This is to avoid
torques on particles in the chain (gravitational torques acting
directly on the particles within a chain are ignored).
(b) A simple realization of the {\sc FPA} model as a rectilinear arrangement
of force chains under tangential loading.
\label{fig:stresspath}}
\end{figure}

Although simplified, we believe that this picture correctly captures some
of the essential physics of force chains. Such
chains are load-bearing structures within the contact network and, in the
simplest approximation of straight chains of uniform orientation these must
have the property described above:  any difference in the forces on two ends
of a path must be balanced by a body force.
Note that if one makes a linear
chain of more than two rigid particles with point contacts, then to avoid
torques, this can indeed support only tangential forces, regardless of the
local friction coefficient between the grains themselves;
see figure~\ref{fig:stresspath}(a).
Force chains should, we believe, be identified (on the average)
with the characteristic
rays of our hyperbolic equations. The {\it mean orientation} of the force
chains is then reflected in a constitutive equation such as {\sc FPA} or
{\sc OSL}.

Our modelling approach thus assumes that the mean
orientation of force chains, in each element the material, is fixed at
burial. (This does not necessarily require that the individual chains are
themselves fixed.) We think it reasonable to assume that the force chains
will not change their average orientations so long as they are able to
support subsequent applied loads. But if a load is applied which they cannot
support (one in which the tangential force difference and body force along a
path do not match) irreversible rearrangement is inevitable 
(Evesque, private communication).
This causes some part of the pile to adopt a
new microtexture and thereby a new constitutive relation. In other words,
{\em incompatible} loadings of this kind must be seen as part of the
construction history of the pile.

\subsection{Fragile matter}

There is a close connection between these ideas and recent work on the
`marginal mechanics' of periodic arrays of identical grains.
(This is considered further in Appendix~\ref{sec:Trollope}.)
The marginal situation is where the (mean) coordination
number of the grains is the minimum required for mechanical integrity;
in two dimensions this is three for frictional and four for frictionless
spheres. (Larger coordination numbers are needed for aspherical grains.)
Indeed, each {\sc OSL} models rigorously describes the continuum mechanics of
a certain ordered array of this kind (see Appendix~\ref{sec:Trollope}).
Marginal packings are exceptional in an obvious sense:
most packings of grains
one can think of do not have this property, and the forces acting on
each grain cannot be found without further information. However, we can
interpret this correspondence between continuum and discrete equations, not
at the level of the packing of individual grains (for which the marginal
coordination state would be hard to explain) but at the level of a granular
skeleton made of force chains.  The {\sc OSL} models (in two dimensions)
can then
be viewed as postulating a simplified, marginally stable geometry of the
skeleton, in which a regular lattice of force chains (bearing tangential
forces only) meet at four-fold coordinated junctions.
(For the {\sc FPA} model, though not in general, this lattice is rectangular.
See figure~\ref{fig:stresspath}(b).)
Such a skeleton leads to hyperbolic equations (or perhaps parabolic ones
if enough disorder is added);
its mechanics are determinate in the absence of a displacement field
specified at the base.

In the present context, fragility arises from the the requirement of
tangential force balance along force chains. If this is violated at the
boundary (within the models as so far defined, even infinitesimally) then
internal rearrangement must occur, causing new force chains to form, so
as to support the load.
It seems reasonable to assume that when
rearrangements are forced upon the system, it responds in an
`overdamped manner' -- that is, the motion ceases as soon as the load is
once again supported. If so, one expects the new state to again be
marginally stable. This suggests a scenario in which the skeleton
evolves dynamically from one fragile state to another. By such a
mechanism, marginally stable packings, although exceptional in the
obvious sense that most packings one can think of are not marginal, may
nonetheless be generic in unconsolidated dry granular matter.
Thornton (1997) reports that, in simulations of frictional spheres,
force chains do rearrange strongly under slight reorientations of the
applied load.

Consider finally a regular lattice of force chains, for
simplicity rectangular (the {\sc FPA} case) which is fragile if the chains can
support only tangential loads. This is the case so long as such paths 
consist of linear chains of rigid particles, meeting at frictional
point contacts: as mentioned above,
the forces on all particles within each chain must then
be colinear, to avoid torques. This imposes the ({\sc FPA}) requirement that
there are no shear forces across a pair of orthogonal planes normal to
the force chains themselves
(see Section~\ref{sec:Strain}\ref{subsec:Anisotropic}).
Suppose now a small degree of particle deformability is allowed
(Cates {\em et al.}, to be published).
This relaxes {\em slightly} the collinearity requirement, but only because
the point contacts are now flattened. The
ratio $\epsilon$ of the maximum transverse load to the normal one will
therefore vanish as some power of the mean deformation. This
yield criterion applies only across two special planes; failure
across others is governed by some smooth yield requirement (such as
the ordinary Coulomb condition: the ratio of the principal stresses
lies between given limits). The granular skeleton just described, which
was fragile in the limit of rigid grains, is now governed by a strongly
anisotropic elastoplastic yield criterion of precisely the kind described
in Section~\ref{sec:Strain}\ref{subsec:Anisotropic}.
The skeleton can support loads that do violate
the tangential balance condition, but only through terms that vanish as
$\epsilon \to 0$. To escape the hyperbolic regime of `fabric dominance',
$\epsilon$ must be significant, which in turn requires significant
particle deformation under the influence of the mean stresses applied.

This indicates how a non-fragile packing of frictional, deformable
rough particles, displaying broadly conventional elastoplastic features
when the deformability is significant, can approach a fragile limit
when the limit of a large modulus is taken at fixed loading. (It does
not, of course, imply that {\em all} packings become fragile in this
limit.) Conversely it shows how a packing that is basically fragile (in
its response to gravity) could nonetheless support very small incremental
deformations, such as sound waves, by an elastic mechanism.
The question of whether sandpiles are
better described as fragile, or as ordinarily elastoplastic, remains open
experimentally. To some extent it may depend on the question being
asked. However, we have argued, on various grounds, that in calculating the
stresses in a pile under gravity a fragile description may lie closer to the
true physics.

\section{Conclusion}
\label{sec:Conclusion}
From the perspective of geotechnical
engineering, the problem of calculating stresses in the
humble sandpile may appear to be of only of marginal importance.
The physicist's view is different: the sandpile is
important, because it is one of the simplest problems in granular
mechanics imaginable. It therefore provides a test-bed for
existing models and, if these show shortcomings, may suggest
ideas for improved physical theories of granular media.

There are, in physics, certain types of problem for which the
fundamental principles or equations are clear, and the difficulty
lies in working out their consequences. An example is the use of
the Navier Stokes equation in studies of (say)
turbulence. The form of the Navier Stokes equation can be deduced
by considering only the symmetries and conservation laws of an isotropic
fluid. Accordingly, its status is not, as sometimes assumed, that
of an approximation based on constitutive hypotheses that happen to be
very accurate for certain materials. Rather, it describes a limiting
behaviour, which all members of a large class of materials (viscoelastic
fluids included) approach with indefinite accuracy in the limit of long
length- and time-scales. 
(We are aware of no theory of elastoplasticity
having remotely similar status.)
There are other types of problem in which the fundamentals are not clear. 
For such problems, the governing equations must first be established, 
before they can be solved. We remain convinced that the static modelling of 
{\em poured assemblies of cohesionless grains under gravity} is of this
second type. This view is not particularly new, either among physicists
(Edwards \& Oakeshott 1989), or among engineers (Gudeshus 1985,1997).

From this perspective, we can see no reason why the
starting points of simple rigid-plastic or elastoplastic
continuum mechanics should offer significant insights into the sandpile
problem. Simple elastoplastic approaches, in particular, give only
one unambiguous physical prediction: that a sandpile supported by a rough
base should have {\em no definite behaviour}.
Experimentalists, who believe themselves to be measuring a definite
result, are likely to be baffled by such predictions. For if, as
these models require, the forces acting at the base of a pile can be
varied at will without causing its static equilibrium to be lost (by
making small elastic displacements at the base), then all the
published `measurements' of such forces must be dismissed as
artefact. An alternative view is that these represent rather haphazard
investigations of some unspecified physical mechanism that does somehow
determine a displacement field at the base of
the pile. (As mentioned previously, basal sag is certainly not an
adequate candidate.) The challenge of whether, for cohesionless poured
sand, such a displacement field can sensibly be defined, remains open.

Given the present state of the data, a conventional elastoplastic
interpretation of the experimental results for sandpiles may remain tenable;
more experiments are urgently required. In the mean time,
a desire to keep using tried-and-tested modelling
strategies until these are demonstrably proven ineffective is quite
understandable. We find it harder to accept the suggestion (Savage 1997) that
anyone who questions the complete generality of traditional elastoplastic
thinking is somehow uneducated.

Our own position is not that elastoplasticity itself is dead,
but we do believe that macroscopic stress propagation in sandpiles is
determined much more by the internal fabric of the material (therefore
the construction history) and much less by boundary conditions, than
{\em simple} elastoplastic models suggest. Reasons for this, based on the
idea of a fragile skeleton of force chains, have been discussed above.
By considering a particular form of yield condition, we have
shown how a fragile model can be matched smoothly onto a relatively
conventional, but strongly anisotropic, elastoplastic theory.
Thus it is possible in principle to have a model which, although strictly
governed by the mixed hyperbolic/elliptic equations of
elastoplasticity, leads to solutions that obey purely hyperbolic equations
everywhere, to within (elastically indeterminate) corrections that are
small in a certain limit. In such a system the results will depend less
and less on boundary conditions, and more and more on fabric, as that
limit is approached.  Moreover, for certain well-defined fragile
packings of frictional grains, the limit is the rigid particle one,
in which the elastic modulus of the grains is taken to infinity at fixed 
loading.

In summary, we have discussed a new class
of models for stress propagation in granular matter. These models assume
local propagation rules for stresses which depend on the construction
history of the material and which lead to hyperbolic differential
equations for the stresses. As such, their physical basis is
substantially different from that of conventional
elastoplastic theory 
(although they may have much more in common with `hypoplastic' models).
Our approach describes a regime of `fragile' behaviour, in which stresses 
are supported by a granular skeleton of force chains that must undergo 
finite internal rearrangement under certain types of infinitesimal load. 
Obviously, such models of granular matter might be incomplete in various ways. 
Specifically we have
discussed a possible crossover to elastic behaviour at very small
incremental loads, and to conventional elastoplasticity at very high mean
stresses (when significant particle deformations arise). However, we believe
that our approach, by capturing at least some of the physics of force
chains, may offer important insights that lie beyond the scope of
conventional elastoplastic or rigid-plastic modelling strategies.
The equivalence between our fragile models and limiting forms of extremely
{\em anisotropic} elastoplasticity, has been pointed out.

\begin{acknowledgments}
We are grateful to S. Edwards, P. Evesque, J.
Goddard, G. Gudehus, J. Jenkins, D. Kolymbas, D. Levine, S. Nagel, S. Savage,
C. Thornton, and T. Witten  for discussions. This research was funded in part
by EPSRC (UK) Grants GR/K56223 and GR/K76733.
\end{acknowledgments}

\begin{appendix}

\section{Laminated elastoplastic cone}
\label{sec:Laminated}
As an alternative to the  `spaceship model',
one might envisage (Fig.~\ref{fig:spaceship}(b)) the creation of a pile
by incremental addition of thin layers of elastoplastic material to its
upper surface (in imitation of an avalanche).
It might then be argued that this thin layer, being under negligible stress,
must be characterized by a zero displacement field (Savage 1998).
On a rough support, one would then expect the displacement at the base to
remain zero as further additions to the pile are made, giving a zero
displacement boundary condition at the base of what has, by now,
presumably become a simple elastoplastic body.

This reasoning is flawed:
the same argument entails that,
at any stage of the pile's construction, the {\em last} layer added is
in a state of zero displacement, not just where it meets the base, but
along its entire length. If so, then not only the base but also the free
surface of the pile is subject to a zero displacement boundary
condition. For a simple elastoplastic cone or wedge, this is incompatible
with the zero stress boundary condition already acting at the free
surface. (Such a body, in effect suspended under gravity from a fixed upper
surface, will exert forces across that surface, as well as across the
supporting base).

The paradox is resolved by noticing that this `laminated elastoplastic'
model in fact involves the addition of thin, stress-free elastoplastic
layers to an already deformed body.
The result will not be a simple elastoplastic continuum, but a body in which
internal stresses and displacements are present even when all body forces
are removed (like a reinforced concrete pillar, or a tennis racket made of
laminated wood) -- Fig.\ref{fig:spaceship}(b). Such a body can, if
carefully designed with a specific loading in mind, satisfy
simultaneously a zero stress and zero displacement (more properly, constant
displacement) boundary condition at any particular surface. These rather
intriguing properties may well be worth investigating further, but they
are still a long way from a realistic description of the construction history
of a sandpile.  In any case it is misleading to suggest (Savage 1998) that
such considerations can justify the adoption of a zero
displacement basal boundary condition within an ordinary ({\em i.e.}, not
pre-strained), isotropic elastoplastic continuum model.

\section{Microscopic force transmission models}
\label{sec:Trollope}
\begin{figure}[]
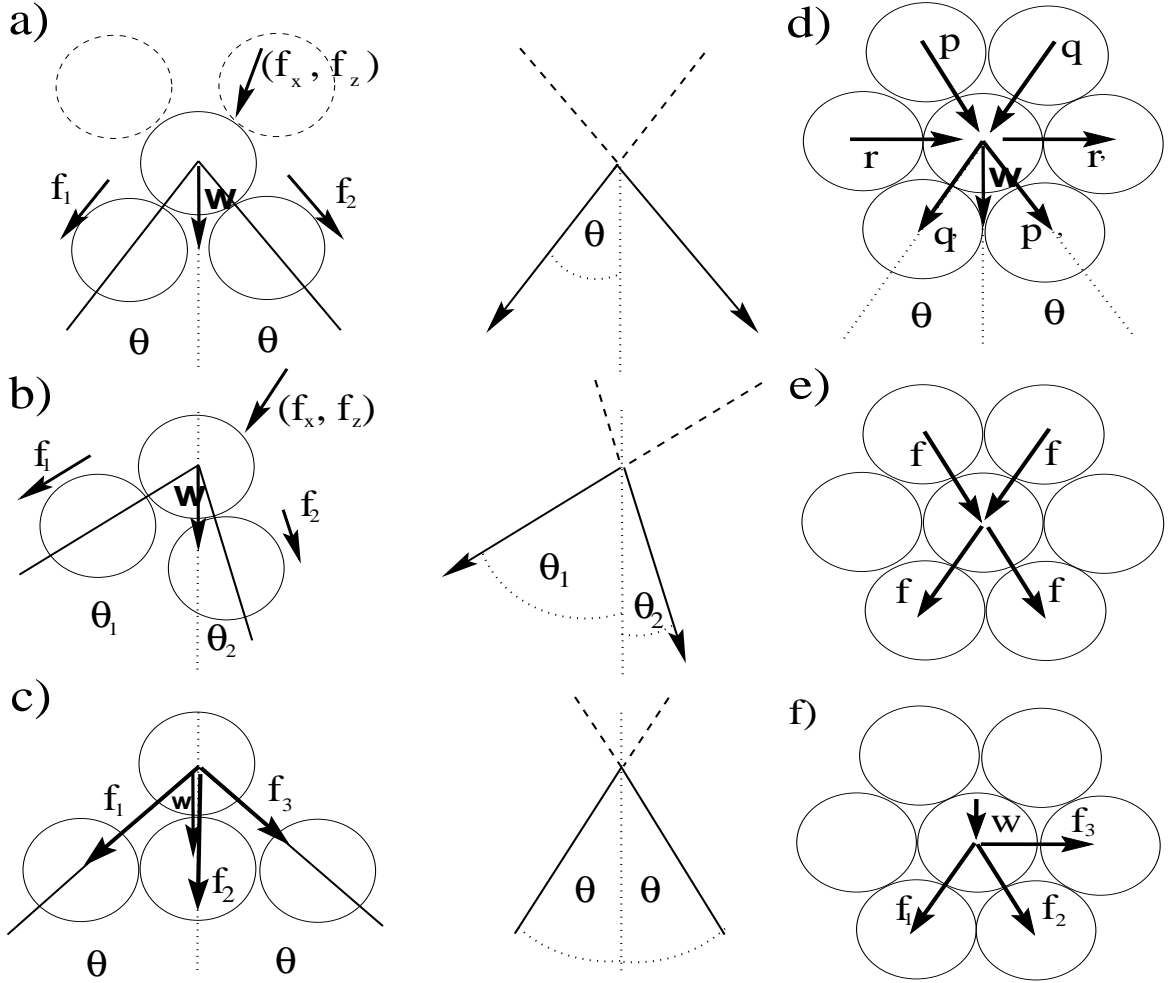

\caption{
(a) Force transfer rules for a simple discrete model (Hong 1993,
Bouchaud {\em et al.} 1995). The forces obey $(f_2-f_1)\sin\theta = f_x$ and
$(f_1+f_2)\cos\theta = f_z+w$ with $w$ that of
gravity. A first order difference equation for $f_x$ is found by writing
$f_x(x,z) = [f_2(x-\Delta x, z-\Delta z) - f_1(x+\Delta x,
z-\Delta z)]\sin\theta$, with $\Delta x = d\cos\theta$ and
$\Delta z = d\sin\theta$, and eliminating $f_{1,2}$ in favour of
$f_{x,z}$ ($d$ is the grain diameter). A similar procedure is then followed
for $f_z$. 
In the continuum limit, the resulting first order differential
equations give the {\sc BCC} model (with $c_0 = \tan\theta$)
with two characteristics (right).
(b) The same, with asymmetric propagation rules, leading to
an asymmetric {\sc OSL} model.
(c) A simple model with three downward
neighbours. The force assignment rule for the middle ray is $f_2 =
\alpha(f_1+f_3)$, where $\alpha$ is some constant.
As shown by Claudin {\em et al.} (1998), the result is still an
{\sc OSL} model (in fact {\sc BCC} with $c_0 = \tan\theta'<\tan\theta$).
(d) In Trollope's model, the outgoing granular forces $(p',q',r')$ 
depend separately on the incoming ones ($p,q,r)$ rather 
than on their vector sum:
$p'-p = w/[c(1+k)]$, $q'-q = wk/[c(1+k)]$ and $r'=r+(1-k)wt/(1+k)$.
Here $w$ is the weight of a grain and $c,t$ denote $\cos\theta, \tan\theta$. 
(e) As a result, for $0<k<1$ a symmetrical extra
loading from two neighbours above whose resultant $2fc$ is
directly downwards, is not equivalent to an increase
(f) in grain weight $w=2fc$.
\label{fig:trollope}}
\end{figure}

Note first that a very large class of discrete models
lead directly to {\sc OSL} models in the continuum limit.
A simple example is defined in Fig.~\ref{fig:trollope}(a). 
As shown by Bouchaud {\em et al.} (1995), this model gives a wave equation 
with two characteristic rays symmetrically arranged about the vertical axis. 
If the symmetry in the stress propagation rules is broken,
an asymmetric {\sc OSL} model arises instead (Fig.~\ref{fig:trollope}(b)).

Secondly, when the continuum
limit of such force-transfer models is taken, one has (in two
dimensions) {\em only two characteristic rays} even if the force
transfer rules involve more than two neighbours in the layer below. 
An example (Claudin {\em et al.} 1998) is shown in Fig.~\ref{fig:trollope}(c).
Broadly speaking, one recovers an {\sc OSL} model, in the continuum limit, 
whenever the forces passed from a grain to its downward (or sideways) 
neighbours obey a
deterministic linear decomposition of the `incident force' $(f_x,f_z)$,
defined as the vector sum of the forces acting from grains in the layer
above, plus the body force on the given grain.

Trollope's model, whose force transfer rules are as shown in
Fig.~\ref{fig:trollope}(d-f), is not a member of this class. (Indeed
it has three characteristic rays in the continuum limit, rather than two.)
This is because {\em the vector sum of the incident forces on a grain is not
taken} before applying a rule to determine the outgoing forces from that
grain; the latter depend {\em
separately} on each of the incident forces. As a description of
hard frictional grains, we consider this unphysical. For, if the grain in
Fig.~\ref{fig:trollope}(d) is subjected to two equal small extra forces $f$
from its two neighbours in the layer above (whose vector sum is  vertical)
the net effect on the outgoing forces should be equivalent to a small
increase in its weight $w = 2f\cos\theta$. Within Trollope's model, this is
not the case. 
Since its propagation rules are linear, any attempt to rectify this feature 
(by taking the vector sum of the forces before propagating these on to the 
next layer) will give an {\sc OSL} model instead.

\end{appendix}

\newpage

\end{document}